\documentclass[aps,pre,reprint]{revtex4-1}
\usepackage{float}
\usepackage{amsmath}
\usepackage{amsfonts}
\usepackage{graphicx}
\usepackage{epstopdf}
\usepackage{epsfig,color}
\usepackage{natbib}

\newcommand{\pd}[2]{\frac{\partial {#1}}{\partial {#2}}}
\newcommand{\td}[2]{\frac{d {#1}}{d {#2}}}

\newcommand{\uh}{\hat{S}}
\newcommand{\ee}{\mathrm{e}}
\newcommand{\ii}{\mathrm{i}}
\newcommand{\wm}{\omega_\mathrm{m}}
\newcommand{\qm}{q_\mathrm{m}}
\newcommand{\cm}{c_\mathrm{m}}
\newcommand{\xim}{\xi_\mathrm{m}}
\newcommand{\Wm}{\Omega_\mathrm{m}}
\newcommand{\eps}{\varepsilon}
\newcommand{\tDelta}{\tilde{\Delta}}

\newcommand{\Dist}[2]{\sqrt{( {#1} )^2 + ( {#2} )^2}}

\begin{document}
\title{Non-reciprocal wave phenomena in spring-mass chains with effective stiffness modulation induced by geometric nonlinearity}

\author{S. P. Wallen\textsuperscript{1} and M. R. Haberman\textsuperscript{1,2}\\
\textsuperscript{1}Applied Research Laboratories, The University of Texas at Austin, Austin, Texas 78758, USA\\
\textsuperscript{2}Department of Mechanical Engineering, The University of Texas at Austin, Austin, Texas 78712-1591, USA
}

\begin{abstract}
	Acoustic non-reciprocity has been shown to enable a plethora of effects analogous to phenomena seen in quantum physics and electromagnetics, such as immunity from back-scattering and unidirectional band gaps, which could lead to the design of direction-dependent acoustic devices. One way to break reciprocity is by spatiotemporally modulating material properties, which breaks parity and time-reversal symmetries. In this work, we present a model for a medium in which a slow, nonlinear deformation modulates the effective material properties for small, overlaid disturbances (often referred to as `small-on-large' propagation). The medium is modeled as a discrete spring-mass chain that undergoes large deformation via prescribed displacements of certain points in the unit cell. A multiple-scale perturbation analysis shows that, for sufficiently slow modulations, the small-scale waves can be described by a linear, monatomic chain with time- and space-dependent on-site stiffness. The modulation depth can be tuned by changing the geometric and stiffness parameters of the unit cell. The accuracy of the small-on-large approximation is demonstrated using direct numerical simulations. 
\end{abstract}

\maketitle

\section{Introduction}

	Acoustic reciprocity is a fundamental physical principle stating that sound propagation between two points is independent of the choice of source and receiver \cite{Strutt1871, Fleury2014,Achenbach2003}, and is generally obeyed except for certain specific scenarios. Breaking acoustic reciprocity allows waves to be tailored differently in different directions, including the possibility of one-way sound propagation \cite{Fleury2014,Haberman2016acoustic,Cummer2016}, and could lead to the design of direction-dependent acoustic devices with the potential to aid in numerous acoustical applications, such as vibration isolation, signal processing, acoustic communication, and energy harvesting.
	
	
	One way to realize acoustic non-reciprocity is by applying a bias that is oddly-symmetric upon time reversal, which has been achieved in moving media \cite{Godin1997,Godin2006}, gyroscopic phononic crystals \cite{Nash2015,Wang2015}, and piezophononic media \cite{Merkel2018}, for example, and effectively establishes `up-stream' and `down-stream' directions for propagating waves. Another means to break reciprocity is nonlinearity, which has been used to create one-way sound propagation via harmonic generation \cite{Boechler2011a,Zhang2016,Bunyan2018}. A third mechanism, which is the subject of the present study, is spatiotemporal modulation of material properties \cite{Cassedy1963,Cassedy1967,Trainiti2016,Vila2017,Nassar2017,Nassar2017a,Nassar2018}. Past studies have demonstrated that effective mechanical properties can be modified using electromagnetic effects in piezoeletric materials \cite{Casadei2012,Chen2014,Chen2016}, magnetorheological elastomers \cite{Danas2012}, and phononic crystals containing electromagnets \cite{Wang2018}. 
	
	Of particular interest to the present work is periodic, wave-like modulation caused by purely mechanical, nonlinear deformation, which has the effect of altering the linearized stiffness and/or mass properties of small disturbances propagating in superposition. This behavior is often referred to as small-on-large propagation, and has been of interest for ultrasonic, non-destructive testing \cite{Renaud2009,Zhang2013} and mechanical metamaterials \cite{Bertoldi2008, Goldsberry2018, Amendola2018}. Non-reciprocal elastic wave propagation via nonlinear deformation has previously been achieved in chains of cylinders in Hertzian contact \cite{Chaunsali2016}, where the effective stiffness of waves propagating transversely to the cylinder axes was modulated by dynamically changing the angles between them. In this case, the nonlinear deformation and overlaid nonreciprocal propagation occurred in degrees of freedom that were naturally decoupled (torsional and longitudinal displacements, respectively); that is, relative torsional displacements between the cylinders altered the effective stiffness for longitudinal waves, but did not directly generate noticeable longitudinal displacements on their own. 
	
	In this work, we present a discrete spring-mass chain model that achieves modulated elastic properties via small-on-large propagation, where the small and large deformations may occur in the same degrees of freedom. The small and large scales are analyzed via multiple-scale perturbation analysis, which provides more specific details about the accuracy of the small-on-large approximation and its range of validity. We find that where the approximation is valid, the linearized equation describing the small-amplitude signal has the same form as a monatomic spring-mass chain with time-dependent on-site stiffness, and this stiffness can be tuned significantly by varying the geometric and stiffness parameters of the unit cell. The linearized chain model is suitable for theoretical analysis using techniques from prior works \cite{Trainiti2016,Vila2017,Attarzadeh2018}. Finally, we demonstrate the effectiveness of the small-on-large approximation by comparing theoretical results to direct numerical simulations of the fully nonlinear equations of motion. The methods presented herein should aid in the design of more complex and/or continuous structures for manipulating mechanical material properties.
	
	\vfill
	
\section{Theoretical Model}

	
	For simplicity, we consider a lumped-element model, as wave propagation with geometric nonlinearity can be easily modeled by systems of ordinary differential equations. While the analysis presented herein could be applied to continuous structures in the future, this would likely require more complex computational techniques (e.g. finite element methods). Thus, a discrete system is sufficient to convey our main results in a straightforward manner.
	
\begin{figure}[t]
		\centering
		\includegraphics[width=0.9\linewidth]{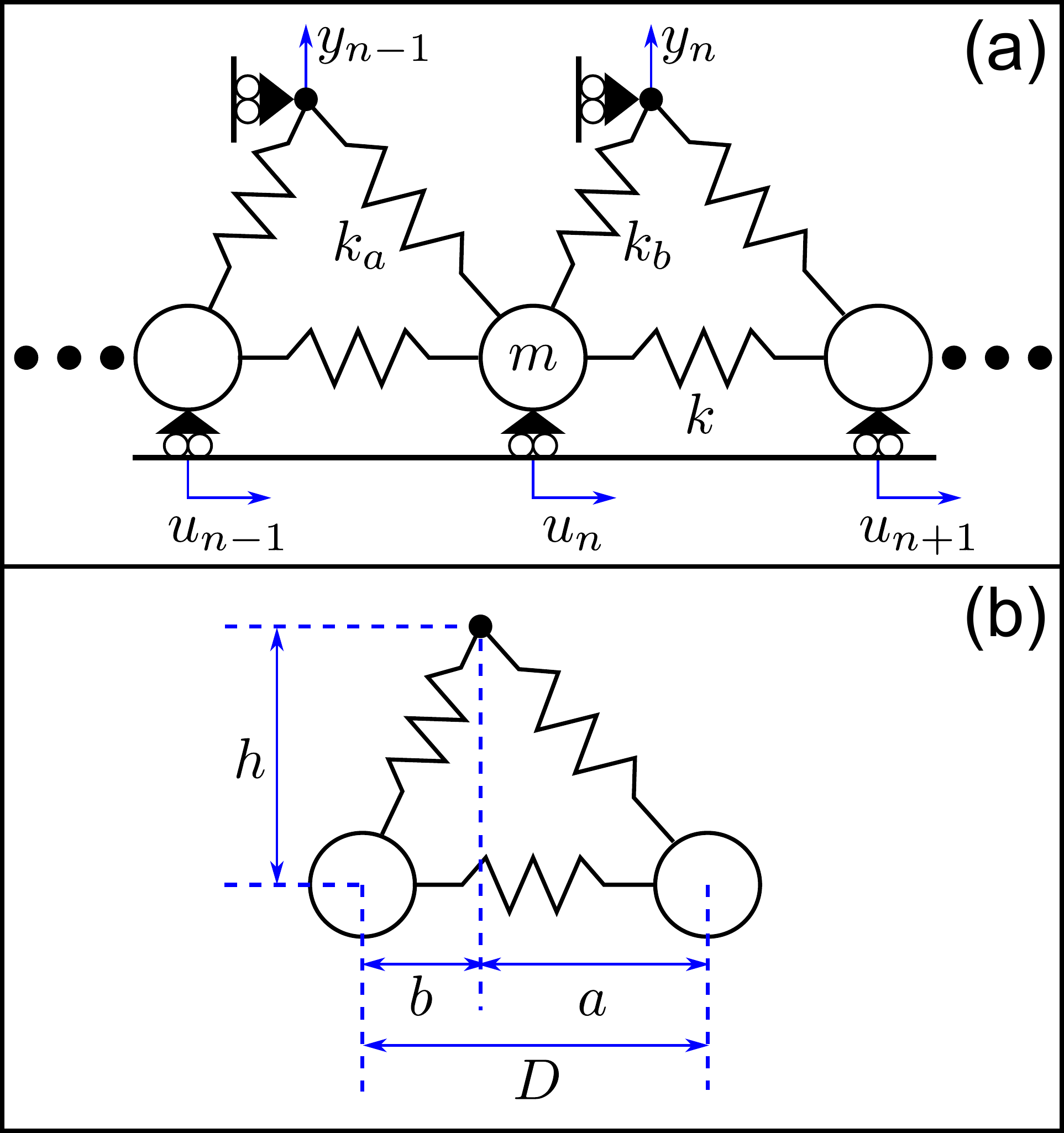}
		\caption{Schematics of representative unit cells of the spring-mass chain. (a) Two unit cells with labeled displacements and lumped element parameters. (b) One unit cell with labeled dimensions.}
		\label{fig:schematic}
	\end{figure}

	Specifically, we consider longitudinal wave propagation in a periodic, monatomic chain of springs and masses with the unit cell shown in Fig. \ref{fig:schematic}(b). The mass $ m $ at the $ n^{th} $ site is coupled to its nearest neighbors and the left and right movable nodes by linear springs with stiffnesses $ k $, $ k_a $, and $ k_b $, respectively, as shown in Fig. \ref{fig:schematic}(a). The horizontal displacement of the $ n^{th} $ mass is denoted $ u_n(t) $ and the vertical displacement of the node to the right is denoted $ y_n(t) $. The modulation is achieved by prescribing $ y_n(t) $ and allowing displacements on the order of the unit cell height $ h $; this causes significant geometric nonlinearity, which alters the incremental stiffness of the structure.
	

	\subsection{Equation of Motion}
	
		To derive the equation of motion of the unit cell, we form the Lagrangian $ \mathcal{L} = \mathcal{T} - \mathcal{V} $, where $ \mathcal{T} $ and $ \mathcal{V} $ represent the kinetic and potential energies, respectively, and are given by the relations
		
			\begin{eqnarray}\label{eq:TV}
				\mathcal{T} &=& \frac{1}{2} m \dot{u}_n^2, \\
				\mathcal{V} &=& \frac{1}{2} k \left(u_{n+1} - u_n \right)^2 + \frac{1}{2} k \left(u_{n} - u_{n-1} \right)^2 \nonumber \\
									&+& \frac{1}{2} k_b \left(\delta_b - l_b \right)^2 + \frac{1}{2} k_a \left(\delta_a - l_a \right)^2.
			\end{eqnarray}
			
		\noindent
		Here, $ \delta_a = \sqrt{ (h+y_{n-1})^2 + (a + u_n)^2 } $ and $ \delta_b = \sqrt{ (h + y_{n})^2 + (b - u_n)^2 } $ are the instantaneous lengths of springs $ k_a $ and $ k_b $, and the corresponding un-stretched spring lengths are given by $ l_a = \sqrt{a^2 + h^2} $ and $ l_b = \sqrt{b^2 + h^2} $. Substituting Eq. \eqref{eq:TV} into Lagrange's equation,
		
			\begin{equation}\label{eq:lagrange}
				\td{}{t} \left[\pd{\mathcal{L}}{\dot{u}_n}\right] - \pd{\mathcal{L}}{u_n} = 0,
			\end{equation}
			
		\noindent
		we obtain the equation of motion for the displacement $ u_n $:
		
			\begin{eqnarray}\label{eq:EOM}
				m \ddot{u}_n &+& k \left( -u_{n-1} + 2 u_n - u_{n+1} \right) \nonumber \\
				&+& k_b \left(\delta_b - l_b\right) \frac{u_n - b}{\delta_b} \nonumber \\
				&+& k_a \left(\delta_a - l_a\right) \frac{u_n + a}{\delta_a} = 0. 
			\end{eqnarray}

	\subsection{Small-on-Large Approximation}

		
		We seek to model small-amplitude `signal' waves in $ u_n $ propagating in the presence of a large, slowly-varying `pump' wave generated by nodal displacements $ y_n $. This is achieved using multiple-scale perturbation analysis \cite{Nayfeh1979}, which allows for separation of the pump and signal wave dynamics.
		
		To begin, we non-dimensionalize Eq. \eqref{eq:EOM} by defining the dimensionless variables $ U_n = u_n/h $, $ Y_n = y_n/h $, and $ T = \omega_0 t $, where $ \omega_0 = \sqrt{k/m} $ is a characteristic frequency. Substitution of these expressions into Eq. \eqref{eq:EOM} gives the following dimensionless equation of motion:
		
		\begin{eqnarray}\label{eq:EOM_ND}
			\frac{d^2 U_n}{dT^2} &+&  \left( -U_{n-1} + 2 U_n - U_{n+1} \right) \nonumber \\
			&+& \kappa_b \left(\Delta_b - \lambda_b \right) \frac{U_n - \beta}{\Delta_b} \nonumber \\
			&+& \kappa_a \left(\Delta_a - \lambda_a \right) \frac{U_n + \alpha}{\Delta_a} = 0,
		\end{eqnarray}
		
		\noindent
		where $ \kappa_{(a,b)} = k_{(a,b)}/k $, $ \lambda_{(a,b)} = l_{(a,b)}/h $, $ \alpha =a/h $, $ \beta = b/h $, $ \Delta_a = \sqrt{ (1+Y_{n-1})^2 + (\alpha + U_n)^2 } $, and $ \Delta_b = \sqrt{ (1 + Y_{n})^2 + (\beta - U_n)^2 } $. Next, we vary the nodal displacements $ Y_n $ in a periodic, wave-like fashion with frequency $ \wm $ and wave number $ \qm $, i.e. 
		
		\begin{equation}
			Y_n = Y_0 \cos{(\theta_n(t))},
			\label{eq:Modulation}
		\end{equation}
		
		\noindent
		where $ \theta_n(t) = \qm nD - \wm t $ is the traveling wave phase and $ Y_0 $ is a constant amplitude. To ensure that $ Y_n $ varies slowly in space and time, we define a small, dimensionless parameter $ \eps << 1 $, and let $ \qm D \propto \eps $ and $ \wm /\omega_0 \propto \eps $. We also define fast and slow time scales $ T_0 = T $ and $ T_1 = \eps T $, respectively, so that 
		
		
		
		\begin{equation}\label{eq:slowDDT}
			\frac{d}{dT} = \frac{d}{dT_0} + \eps \frac{d}{dT_1},
		\end{equation}
		
		\noindent
		as well as the slow, dimensionless spatial variable $ X_1 = \eps n $.	To solve Eq. \eqref{eq:EOM_ND}, we seek a solution of the form
		
		\begin{eqnarray}
			U_n &=& P(D \eps n, T_1) + \eps S_n(T_0) \nonumber\\
				   &=& P(D X_1, T_1) + \eps S_n(T_0), \label{eq:Ansatz1}\\
			U_{n \pm 1} &=& P(D \eps(n \pm 1), T_1) + \eps S_{n \pm 1}(T_0) \nonumber \\
			                    &=& P(D(X_1 \pm \eps), T_1) + \eps S_{n \pm 1}(T_0), \label{eq:Ansatz2}
		\end{eqnarray}
		
		\noindent
		where $ P(X_1, T_1) $ and $ S_n(T_0) $ have the roles of pump and signal waves, respectively, and depend on separate time scales. In choosing this trial solution, we have assumed \textit{a priori} that $ P $ varies slowly, with the same length and time scales as $ Y_n $. To ensure the validity of the approximate solutions that follow, this assumption must be checked for specific parameter sets after the complete solution is obtained.
		
		Substituting Eqs. \eqref{eq:slowDDT} - \eqref{eq:Ansatz2} into Eq. \eqref{eq:EOM_ND} results in the following relationship:
		
		\begin{eqnarray}
			&\eps &\left[ \td{^2S_n}{T_0^2} - S_{n+1} + 2S_n - S_{n-1} \right] \nonumber \\
		+ &\eps^2 &\left[ \pd{^2P}{T_1^2} - \pd{^2 P}{X_1^2}\right] \nonumber \\
		+&\kappa_b &\frac{\Dist{1+Y_n}{\beta - P - \eps S_n} - \lambda_b}{\Dist{1+Y_n}{\beta - P - \eps S_n}} (P + \eps S_n - \beta) \nonumber \\
		+&\kappa_a &\frac{\Dist{1+Y_{n-1}}{\alpha + P + \eps S_n} - \lambda_a}{\Dist{1+Y_{n-1}}{\alpha + P + \eps S_n}} (P + \eps S_n +\alpha) \nonumber \\
		 = &0&, \label{eq:Expansion1}
		\end{eqnarray}
		
		\noindent
		where we have used the Taylor expansion 
		
		\begin{equation}\label{key}
			P(D(X_1 \pm \eps), T_1) \approx P(DX_1,T_1) \pm \eps \pd{P}{X_1} + \frac{\eps^2}{2} \pd{^2P}{X_1^2}.
		\end{equation}
		
		\noindent
		Finally, we expand the nonlinear terms of Eq. \eqref{eq:Expansion1} as Taylor series in $ \eps S_n $ about $ \eps S_n = 0 $, collect terms proportional to each power of $ \eps $, and find the following equations describing the dynamics of the pump and signal waves (accurate to $ \mathcal{O}(\eps) $):

		\hfill \\
		
		\noindent
		$ \mathcal{O}(1) $:
		\begin{eqnarray}
			\kappa_b(P_n-\beta)\frac{\tDelta_b-\lambda_b}{\tDelta_b}
			+ \kappa_a(P_n+\alpha)\frac{\tDelta_a-\lambda_a}{\tDelta_a} = 0, \label{eq:Order1}
		\end{eqnarray}
		
		\noindent
		$ \mathcal{O}(\eps) $:
		
		\begin{eqnarray}
			\frac{d^2 S_n}{dT_0^2} &+&  \left( -S_{n-1} + 2 S_n - S_{n+1} \right) + K_n(P_n)S_n = 0, \nonumber \\
			&& \label{eq:OrderEps}\\
			K_n(P_n) &=& \kappa_b \frac{\tDelta_b^2(\tDelta_b-\lambda_b) + \lambda_b(P_n - \beta)^2}{\tDelta_b^3} \nonumber \\
			&+& \kappa_a \frac{\tDelta_a^2(\tDelta_a-\lambda_a) + \lambda_a(P_n + \alpha)^2}{\tDelta_a^3} \label{eq:Kn},
		\end{eqnarray}
		
		\noindent
		where $ \tDelta_a = \sqrt{(1+Y_{n-1})^2 + (\alpha+P_n)^2} $ and $ \tDelta_b = \sqrt{(1+Y_n)^2 + (\beta-P_n)^2} $.

	\subsection{Effective Linear Chain}
\begin{figure}[t]
			\centering
			\includegraphics[width=0.7\linewidth]{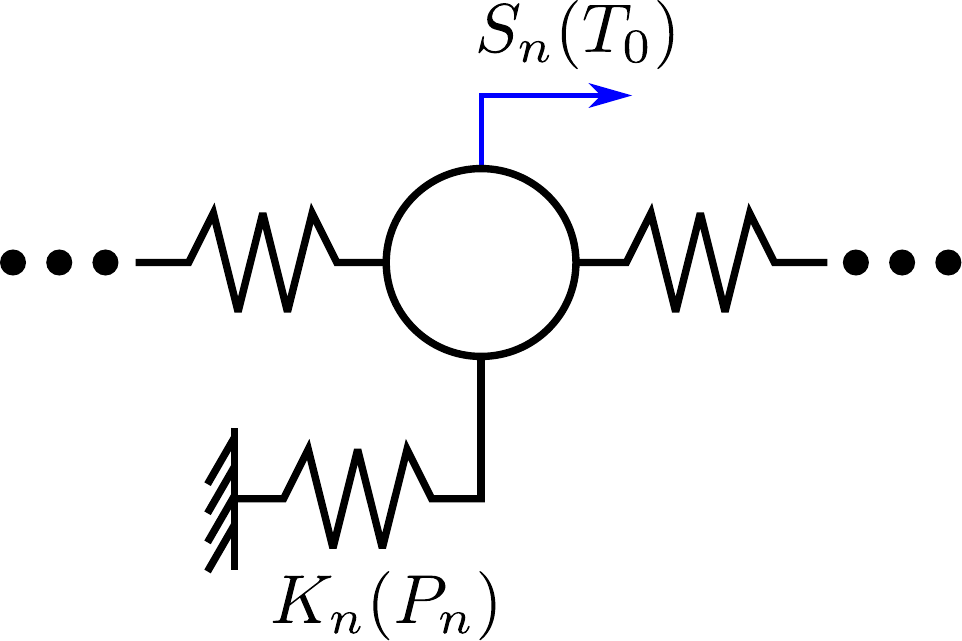}
			\caption{Schematic of one unit cell of a linear, time-dependent, monatomic chain describing the dynamics of the signal wave $ S_n(T_0) $.}
			\label{fig:linSchematic}
		\end{figure}

		Equation \eqref{eq:OrderEps} has the form of a linear, monatomic spring-mass chain with space- and time-dependent on-site stiffness $ K_n $, as shown schematically in Fig. \ref{fig:linSchematic}. As is evident from Eq. \eqref{eq:Kn}, this on-site stiffness depends on the pump wave $ P_n $, which can calculated for prescribed $ Y_n $ using Eq. \eqref{eq:Order1}. 
		
\begin{figure}[h]
			\centering
			\includegraphics[width=\linewidth]{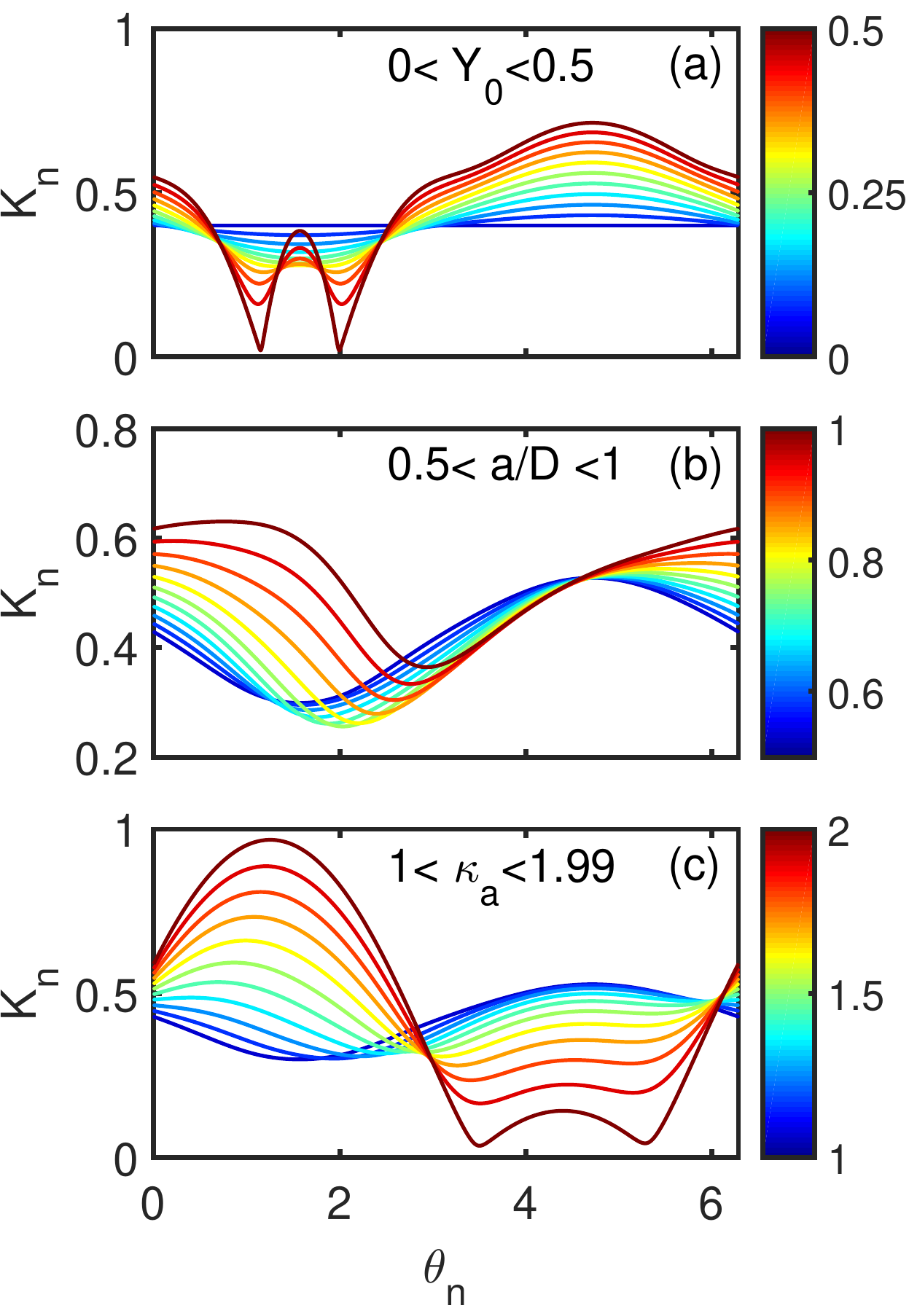}
			\caption{Effective on-site stiffness $ K_n $ as a function of traveling wave phase $ \theta_n $, for (a) variation of nodal displacement amplitude $ Y_0 $ in a symmetric unit cell; (b) variation of unit cell shape $ a/D $ with $ D = 1 $, $ \kappa_a = \kappa_b = 1 $, and $ Y_0 = 0.2 $; and (c) variation of stiffness $ \kappa_a $ with $ \kappa_a+\kappa_b = 2 $, $ a/D = 0.5 $, and $ Y_0 = 0.2 $. Colormaps indicate the value of the varied parameter for each curve.}
			\label{fig:effStiffness}
		\end{figure}

		By varying the geometric and stiffness parameters of the unit cell, we find a wide range of tunability in $ K_n $. The variation of $ K_n $ over one period of the traveling wave phase $ \theta_n $ is shown in Fig. \ref{fig:effStiffness}, where we have numerically solved for $ P_n $ using Eq. \eqref{eq:Order1} with $ Y_n $ given by Eq. \eqref{eq:Modulation}, and substituted the results into Eq. \eqref{eq:Kn}. Cases of varied modulation amplitude, unit cell shape, and unit cell stiffness distribution are shown in Fig. \ref{fig:effStiffness}(a), Fig. \ref{fig:effStiffness}(b), and Fig. \ref{fig:effStiffness}(c), respectively. We note that care must be taken to validate the assumption of a slowly-varying pump wave. Specifically, while the prescribed modulation $ Y_n $ can always be made to vary slowly by enforcing $ \qm D \propto \eps $, the pump wave $ P_n $ and stiffness $ K_n $ depend nonlinearly on $ Y_n $, and therefore may contain harmonics that do not vary slowly in comparison to the signal wave. In situations where significant harmonics are present (e.g. the dark red curves in all three panels of Fig. \ref{fig:effStiffness}, which have sharp changes as a function of the traveling wave phase), $ \qm $ and $ \wm $ must be made sufficiently small for the harmonics to be considered slow as well.
		
		Finally, we remark that strong stiffness modulations can be found in the presence of mechanical instabilities, i.e. buckling. The dark red curves in Fig. \ref{fig:effStiffness}(a) and Fig. \ref{fig:effStiffness}(c) approach zero stiffness at a few points in the cycle due to proximity to a point of instability.  While the parameters considered in this work were chosen to keep the effective stiffness everywhere positive, the points of instability can be reached by further increasing the respective parameter variations. These instabilities, while interesting in their own right (see \cite{Kochmann2017} and references therein), would violate the assumptions of a slowly-varying pump wave and are outside the scope of the present study. Nevertheless, buckling structures may be useful in future studies to achieve strong and highly tunable modulations by operating near, but not fully reaching, instability.

\section{Non-reciprocal Traveling Waves}

%
	
	\subsection{Periodic Traveling Wave Solutions}
	
		To find periodic, traveling wave solutions of Eq. \eqref{eq:OrderEps}, we use a Bloch wave expansion approach similar to the one used in Ref. \cite{Trainiti2016}. We assume solutions of the form
		
		\begin{equation}\label{eq:ansatz}
			S_n = \ee^{\ii \left(\xi n-\Omega T_0 \right)} \sum_{j=-\infty}^{+\infty} \uh_j \ee^{\ii j\Theta_n \left(T_0 \right)},
		\end{equation}
	
		\noindent
		where $ \xi $ and $ \Omega $ are a dimensionless wave number and frequency, respectively, and $ \Theta_n(T_0) = \xim n - \Wm T_0 $ is the phase of the traveling wave modulation in terms of the dimensionless parameters $ \xim = \qm D $ and $ \Wm = \wm/\omega_0$. Equation \eqref{eq:ansatz} is an infinite sum of plane waves with wave numbers $ \xi \pm j \xim $ and frequencies $ \Omega \pm j \Wm $, with corresponding complex amplitude coefficients $ \uh_j $. Thus, when the on-site stiffness is modulated in space and time, the $ (\xi, \Omega) $ spectrum is not a `dispersion relation' in the traditional sense, because the presence of one plane wave necessitates the existence of others. The modulated stiffness $ K_n $ must also be represented as a summation of plane waves:
		
		\begin{equation}\label{eq:Kfourier}
			K_n(\Theta_n(T_0)) = \sum_{p=-\infty}^{+\infty} \hat{K}_p \ee^{\ii p\Theta_n(T_0)},
		\end{equation}
		
		\noindent
		where the amplitude coefficients $ \hat{K}_p $ are calculated from the well-known formula for a complex Fourier series:
		
		\begin{eqnarray}\label{Kcoeffs}
			\hat{K}_{\pm p} = \frac{1}{2 \pi} \int_{0}^{2 \pi} K(\Theta_n)\ee^{-\ii(\pm p)\Theta_n} d\Theta_n.
		\end{eqnarray}
		
		\noindent
		Substituting Eqs. \eqref{eq:ansatz} - \eqref{Kcoeffs} into Eq. \eqref{eq:OrderEps} and utilizing the orthogonality of the complex exponential functions to eliminate one of the summations, we find the hierarchy of equations
		
		\begin{eqnarray}\label{eq:Spectrum}
			-\left( \Omega + l \Wm \right)^2 \uh_l &+& 4 \sin^2 \left(\frac{\xi + l \xim}{2}\right) \uh_l \nonumber \\
			&+& \sum_{j=-\infty}^{+\infty} \hat{K}_{l-j} \uh_j = 0,
		\end{eqnarray}
		
		\noindent
		where the free index $ l = p+j $ arises from orthogonality. To solve Eq. \eqref{eq:Spectrum}, we must truncate the infinite series in the final term of Eq. \eqref{eq:Spectrum} at some value $ \pm J $. Then, the indicies $ j $ and $ l $ take on integer values $ [-J, -(J-1), ... , 0 , J-1, J] $, and for fixed wavenumber $ \xi $, Eq. \eqref{eq:Spectrum} may be cast as a quadratic eigenvalue problem with eigenvalues $ \Omega $ and eigenvectors $ [\uh_{-J}, ... 0, ..., \uh_J]^T $ \cite{Trainiti2016}. The quadratic eigenvalue problem can be solved numerically using standard computational software (in this work, we have used the {\it polyeig} function in MATLAB).
		
\begin{figure}[h]
			\centering
			\includegraphics[width=\linewidth]{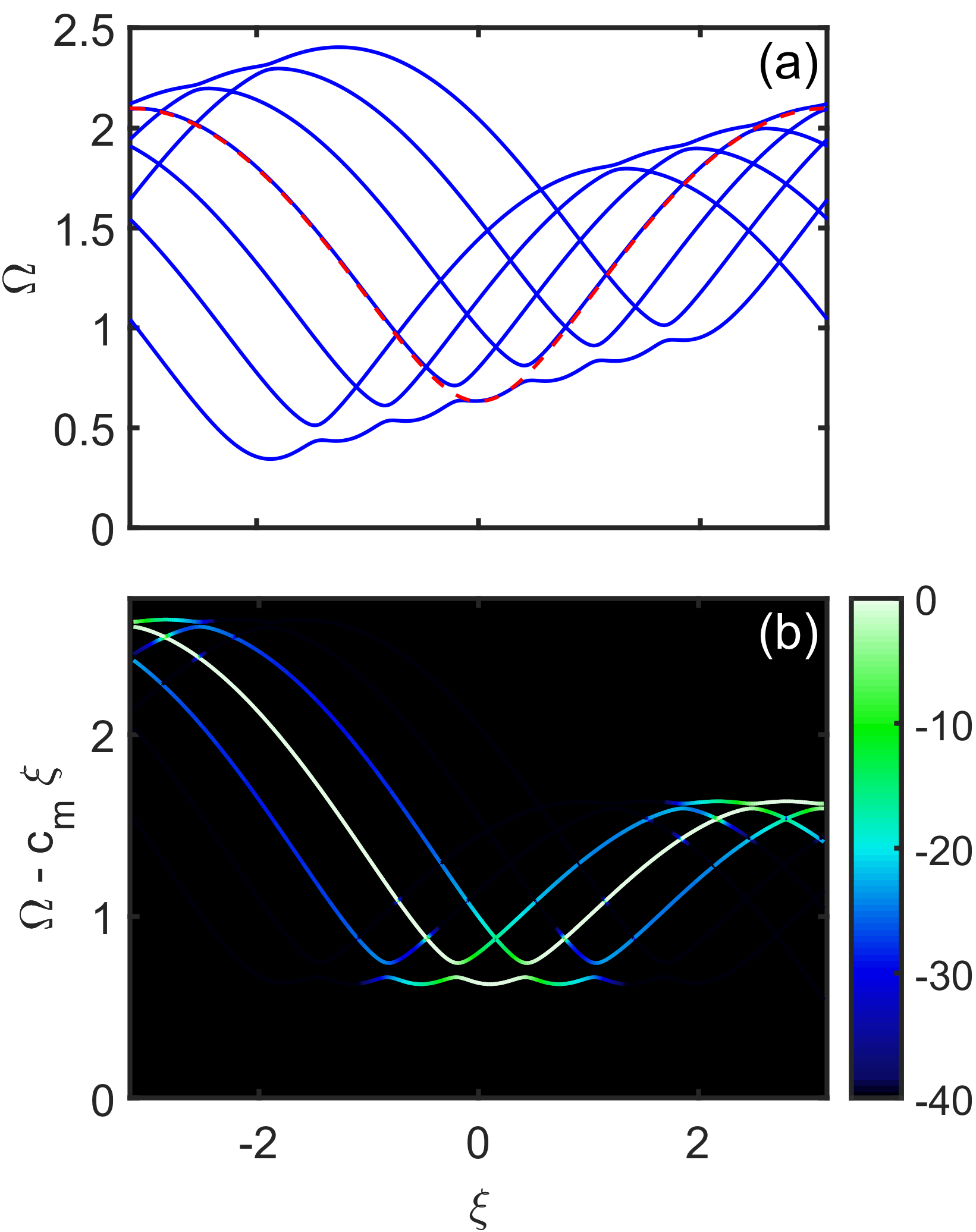}
			\caption{Theoretical band structure for parameters $ \alpha = \beta = \kappa_a = \kappa_b = 1 $, $ Y_0 = 0.2 $, $ \xim = 2\pi/10 $, and $ \Wm = 0.1 $. (a) Frequency-wave number spectra of the modulated (blue solid curves) and non-modulated (red dashed curves) linear chains. (b) Frequency-wave number spectrum of the modulated linear chain with amplitude visible and tilt removed (color map: normalized eigenvector components $ \uh_l / |\uh| $, in decibels).}
			\label{fig:nonRepBands}
		\end{figure}

		Example solutions of Eq. \eqref{eq:Spectrum} for $ J = 3 $ and the parameter values $ \alpha = \beta = \kappa_a = \kappa_b = 1 $, $ Y_0 = 0.2 $, $ \xim = 2\pi/10 $, and $ \Wm = 0.1 $, as well as the solution for an un-modulated chain with equivalent mean on-site stiffness, are shown in Fig. \ref{fig:nonRepBands}(a), where non-reciprocity is evident from asymmetry about the $ \xi=0 $ axis.  This $ (\xi, \Omega) $ spectrum contains $ 2J+1 $ bands, which appear similar to the single band of the un-modulated chain, tiled along a line of slope $ \cm = \Wm/\xim $ \cite{Nassar2018}. However, Fig. \ref{fig:nonRepBands}(a) does not show the relative amplitudes of the plane waves in the solution, nor does it indicate which plane waves belong to each mode; thus, a more intuitive representation of the spectrum can be found by 1) coloring each $ (\xi, \Omega) $ point according to the magnitude of its eigenvector component, and 2) removing the `tilt' due to the temporal modulation, i.e. plotting $ \Omega - \cm \xi $ on the ordinate axis instead of $ \Omega. $ This representation is shown in Fig. \ref{fig:nonRepBands}(b), from which it can be seen that most of the energy in each mode (a `mode' now corresponding to points on any horizontal line) is concentrated near the un-modulated branch.
	
\section{Numerical Results}

	\begin{figure}[h]
		\centering
		\includegraphics[width=\linewidth]{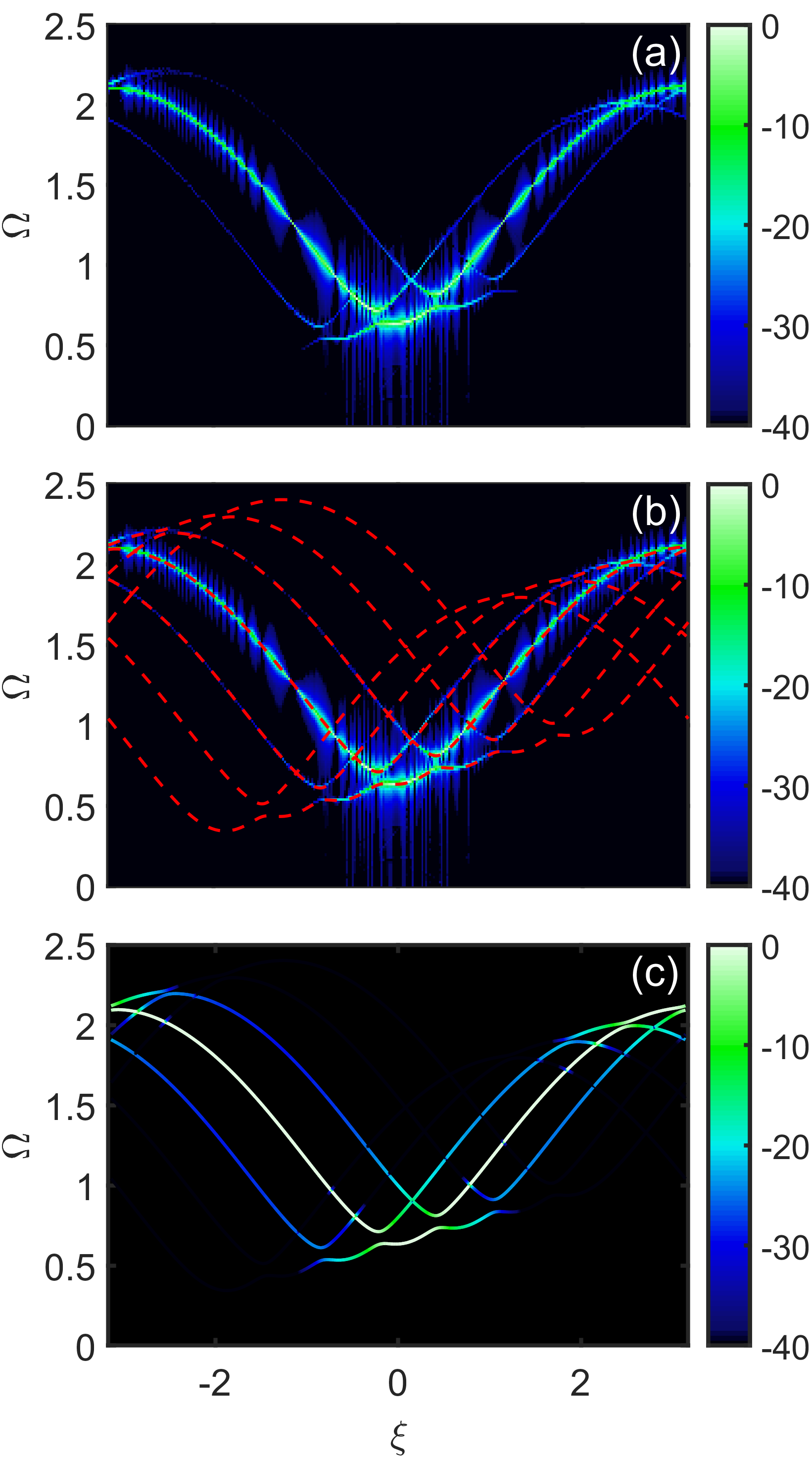}
		\caption{(a) Two-dimensional Fourier transform of the simulated spatiotemporal data (colormap: magnitude of each Fourier coefficient, normalized by the maximum magnitude, in decibels). (b) Fourier transform data from panel (a) with theoretical curves overlaid. (c) Theoretical amplitude-colored spectrum (same data as in Fig. \ref{fig:nonRepBands}(b), shown with tilt restored). }
		\label{fig:FFT}
	\end{figure}

	To demonstrate the effectiveness of the linearized model developed using the small-on-large approximation, we simulate Eq. \eqref{eq:EOM_ND} (the fully nonlinear equations of motion) with the same parameters used in the previous section, using a standard fourth-order Runge-Kutta scheme, and compare the results to the linear theoretical model. The simulation is performed with a chain length of 600 masses and a dimensionless time step of $ \Delta T_0 = 0.01 $, for a duration of $ 8 \times 10^4 $ time steps. With the pump wave present, a broad-band signal wave is imparted to the chain by giving one of the masses an initial dimensionless velocity of 0.01. To isolate the signal wave, we perform a second simulation with the pump wave only, and subtract the result from the first simulation. We note that while the subtraction of the pump wave does not yield the signal wave exactly (due to nonlinearity), it gives an accurate representation of the signal wave as long as the small-on-large approximation is valid. Finally, we perform a two-dimensional Fast Fourier Transform on the resulting spatiotemporal data, giving a numerical frequency-wave number spectrum, which is shown in Fig. \ref{fig:FFT}(a). For comparison, we repeat this spectrum with overlaid theoretical curves from Fig. \ref{fig:nonRepBands}(a) (modulated case) in Fig. \ref{fig:FFT}(b), and include the theoretical, color-coded band structure (i.e. the data from Fig. \ref{fig:nonRepBands}(b) with the tilt restored) in Fig. \ref{fig:FFT}(c). Overall, we find excellent agreement between the fully nonlinear, numerical results and the linearized, small-on-large theoretical model.

\section{Conclusion}

	We have developed a model for non-reciprocal elastic wave propagation via modulated stiffness in a discrete spring-mass chain, where the modulation is achieved by applying a slow, nonlinear deformation that alters the effective on-site stiffness in a quasi-static manner. By applying multiple-scale perturbation analysis, we have shown that in the presence of a sufficiently slow, nonlinear deformation (the `pump' wave), a small-amplitude disturbance (the `signal' wave) can be accurately modeled by a linear spring-mass chain with time-dependent properties. This effective linear chain model may be analyzed using existing methods from recent works. By tuning the material and geometric parameters of our unit cell, we have shown that the modulation depth of the effective stiffness is highly variable. In particular, it can be made large (e.g. on the order of its mean value) when operating near mechanical instabilities. Finally, we have demonstrated the effectiveness of the linearized model by comparing the theoretical results to direct numerical simulations of the fully-nonlinear chain, and found excellent agreement.
	
	Opportunities for future studies include applying similar analyses to chains in which the effective inter-site stiffness is also modulated by nonlinearity (e.g. by allowing the displacement-controlled nodes to move horizontally) and to continuous structures (for example, negative-stiffness honeycomb lattices, which have been shown to exhibit large effective property changes under an applied strain \cite{Correa2015,Goldsberry2018}). Methods for optimizing these structures to obtain a targeted non-reciprocal response should be explored. It would also be of interest to revisit the application of homogenization techniques to modulated media in the context of Willis constitutive equations \cite{Willis1981,Nassar2015willis,Muhlestein2017,Sieck2017}, which have been discussed briefly in some prior specific cases \cite{Nassar2017a,Torrent2018}. Finally, experimental realizations of mechanically-modulated structures are also within reach, as highly deformable elastic lattices can be fabricated using additive manufacturing techniques \cite{Raney2015,Bertoldi2017,Kochmann2017,Correa2015}.

\section*{Acknowledgments}

This work supported by National Science Foundation EFRI award no. 1641078 and the postdoctoral fellowship program at Applied Research Laboratories at The University of Texas at Austin.

\vfill


\bibliographystyle{apsrev4-1}
\bibliography{Mendeley_UTEFRI}

\end{document}